\documentclass[12pt]{revtex4}

\usepackage{graphicx}
\usepackage{hyperref}

\begin{document}

\title{Experimental evidence of spontaneous symmetry breaking in
intracavity type-II second harmonic generation with triple
resonance}

\author{Laurent Longchambon}\affiliation{Laboratoire
Kastler Brossel, UPMC, Case 74, 4 Place Jussieu, 75252 Paris cedex
05, France}
\author{Nicolas Treps}\affiliation{Laboratoire
Kastler Brossel, UPMC, Case 74, 4 Place Jussieu, 75252 Paris cedex
05, France}
\author{Thomas Coudreau}\email{coudreau@spectro.jussieu.fr}\affiliation{Laboratoire
Kastler Brossel, UPMC, Case 74, 4 Place Jussieu, 75252 Paris cedex
05, France}
\author{Julien Laurat}\affiliation{Laboratoire
Kastler Brossel, UPMC, Case 74, 4 Place Jussieu, 75252 Paris cedex
05, France}
\author{Claude Fabre } \affiliation{Laboratoire
Kastler Brossel, UPMC, Case 74, 4 Place Jussieu, 75252 Paris cedex
05, France}

\date{\today}%
%
%
\begin{abstract}
We describe an experiment showing a spontaneous symmetry breaking
phenomenon between the intensities of the ordinary and
extraordinary components of the fundamental field in intracavity
type-II harmonic generation. It is based on a triply resonant dual
cavity containing a type II phase matched $\chi^{(2)}$ crystal
pumped at the fundamental frequency $\omega_0$. The pump beam
generates in the cavity a second harmonic mode at frequency
$2\omega_0$ which acts as a pump for frequency degenerate type II
parametric down conversion. Under operating conditions which are
precisely symmetric with respect to the ordinary and extraordinary
components of the fundamental wave, we have observed a breaking of
the symmetry on the intensities of these two waves in agreement
with the theoretical predictions.
\end{abstract}

\maketitle

Triply Resonant Optical Parametric Oscillators are well-known to
have a rich dynamical behavior showing in particular bistability,
self-oscillation and chaos \cite{lugiato88,lefranc04}. We study
here a similar system consisting in a type-II phase matched
crystal placed inside a triply resonant cavity but where the
pumping is made at the fundamental frequency $\omega_0$, not at
the harmonic frequency $2\omega_0$. In this case, both second
harmonic generation and parametric down-conversion simultaneously
occur. The pump wave is sent at $45^\circ$ of the crystal neutral
axes and is frequency doubled. The produced harmonic wave acts
then as a pump for an OPO inside the same cavity and using the
same crystal. This system has been widely studied theoretically
for its classical and quantum properties
\cite{eschmann93,ou94,marte95,jack96,peschel98}. The system is
\emph{a priori} symmetric : the ordinary and extraordinary
fundamental waves intensities are in general equal at the output
of an OPO. However, a spontaneous symmetry breaking phenomenon is
predicted to occur under some circumstances~: the intracavity
signal and idler intensities may be different. In this letter, we
report what is to our knowledge the first experimental
demonstration of this phenomenon.

\begin{figure}
\centerline{\includegraphics[width=\columnwidth]{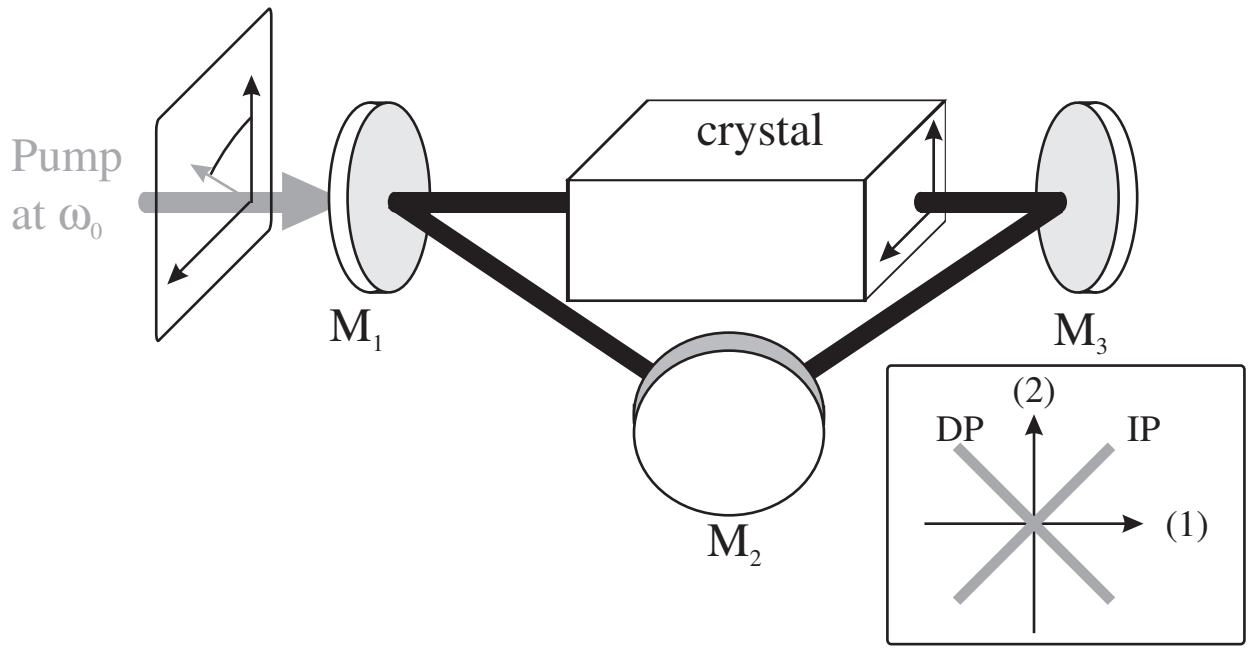}}
\caption{\label{fig:ringcav}Scheme of the ring cavity. M$_1$ :
pump coupling mirror; M$_{2,3}$ : highly reflective mirrors at
$\omega_0$ and $2\omega_0$. In the box : input polarization (IP)
and down-converted polarization (DP) with the crystal neutral axes
(1) and (2).}
\end{figure}

We consider a ring cavity (figure (\ref{fig:ringcav})) containing
a $\chi^{(2)}$ crystal with type-II phase matching.
This cavity is pumped with a beam at frequency $\omega_0$
polarized at $45^\circ$ of the crystal neutral axes, (1) and (2).

Pumped by the intracavity second harmonic field, parametric
oscillation takes place on the down-converted fields frequencies
$\omega_1$ and $\omega_2$ (with $\omega_1+\omega_2=2\omega_0$)
which minimize the threshold \cite{fabre97}. The non frequency
degenerate case has been considered both theoretically and
experimentally \cite{marte94,schiller97,white97} and spontaneous
symmetry breaking is not predicted in this case. Here, we suppose
that the the system is operated at frequency degeneracy
($\omega_1=\omega_2=\omega_0$)~: the down-converted fields are at
the frequency of the input field and interfere with it. We denote
$A_{1,2}$ the intracavity field enveloppes at frequency $\omega_0$
and $A_0$ the intracavity field enveloppe at frequency
$2\omega_0$.  The corresponding detunings are denoted
\begin{equation}
\Delta_{1,2}= \frac{\omega_0}{c}\left( n_{1,2} l + L \right)
[2\pi] \mbox{~and~} \Delta_0= \frac{2\omega_0}{c}\left( n_0 l + L
\right) [2\pi].
\end{equation}
where $n_{0,1,2}$ are the crystal indices of refraction, $l$ is
the crystal length and $L$ the free propagation length inside the
cavity.  When the optical length of the cavity is adjusted so that
all fields are close to resonance and the intracavity losses
small, the normalized equations for the normalized intracavity
field enveloppes can be written~:
\begin{eqnarray}
(\gamma^\prime-i \Delta_1) A_1 &=& A_0 A_2^\ast + \sqrt{2 \gamma}
\frac{A_{in}}{\sqrt 2} \label{eq:A1} \\
(\gamma^\prime-i \Delta_2) A_2 &=& A_0 A_1^\ast + \sqrt{2 \gamma}
\frac{A_{in}}{\sqrt 2} \label{eq:A2}\\
(\gamma_0-i \Delta_0) A_0 &=& - A_1 A_2 \label{eq:A0}
\end{eqnarray}
where $A_{in}$ is the pump field, $\gamma^\prime$ and $\gamma_0$
correspond respectively to the round trip amplitude losses for the
pump and second harmonic wave, while $\sqrt{2\gamma}$ is the
transmission of the coupling mirror $M_1$ at frequency
$\omega_0$). In a standard OPO, the system oscillates on the
frequency pairs which verify
\begin{equation}
\Delta_1=\Delta_2= \Delta. \label{eq:deltas}
\end{equation}
In our case, the frequencies are fixed by the injected field but
we assume, which is experimentally realistic, that the system
parameters can be adjusted so that relation (\ref{eq:deltas}) is
verified. Equation (\ref{eq:A0}) leads to an expression for $A_0$
as a function of $A_1$ and $A_2$ which can be re-injected in
equations (\ref{eq:A1}) and (\ref{eq:A2}). One obtains then a
system of two nonlinear coupled equations for $A_1$ and $A_2$.
From this system, one can obtain the following relation~:
\begin{equation}
\left( (\gamma^{\prime2} + \Delta^2) - \frac{1}{\gamma_0^2 +
\Delta_0^2} I_1 I_2\right)(I_1 - I_2) = 0 \label{eq:fin}
\end{equation}
where $I_i=|A_i|^2$ is the intensity of the corresponding field.
This equation has two solutions, one symmetric ($I_1=I_2 = I$) and
another dissymmetric.

\begin{figure}[h]
\begin{center}
\includegraphics[width=\columnwidth]{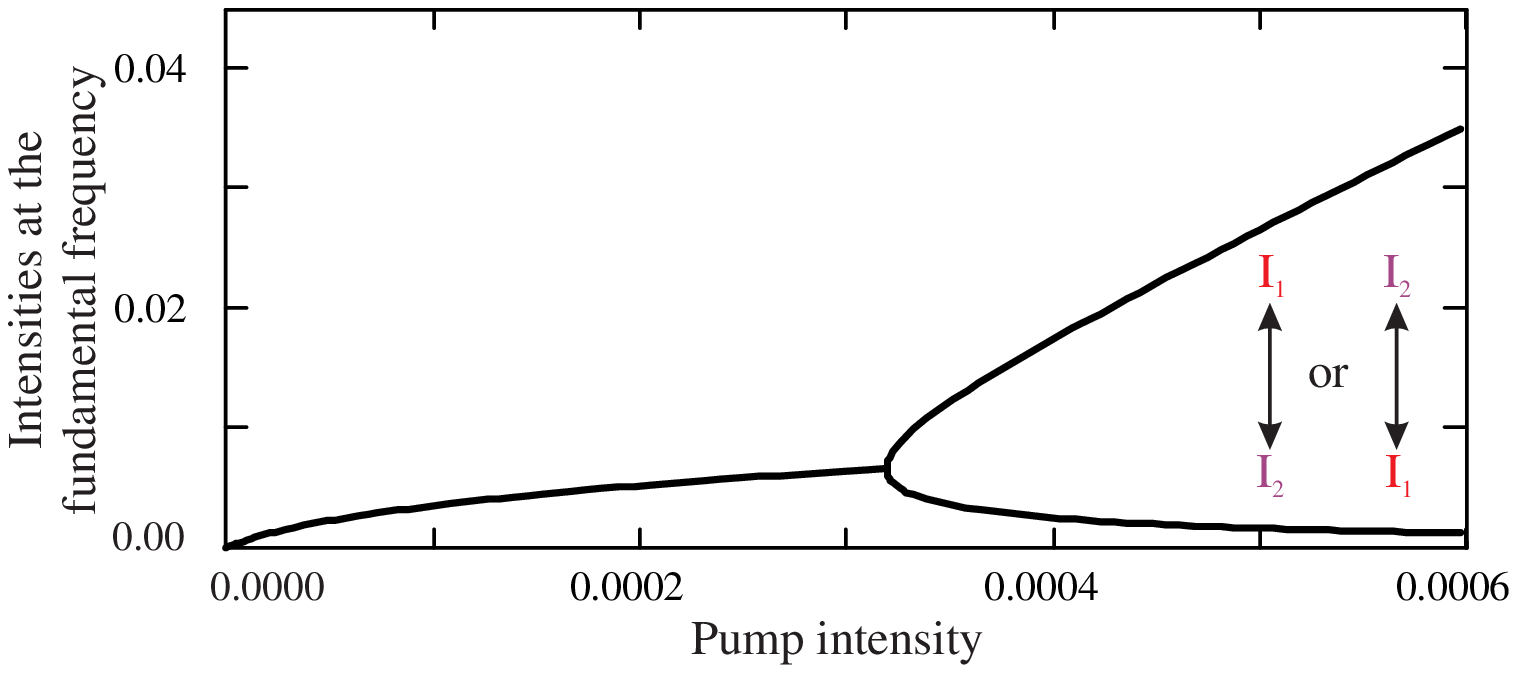}\\
\includegraphics[width=\columnwidth]{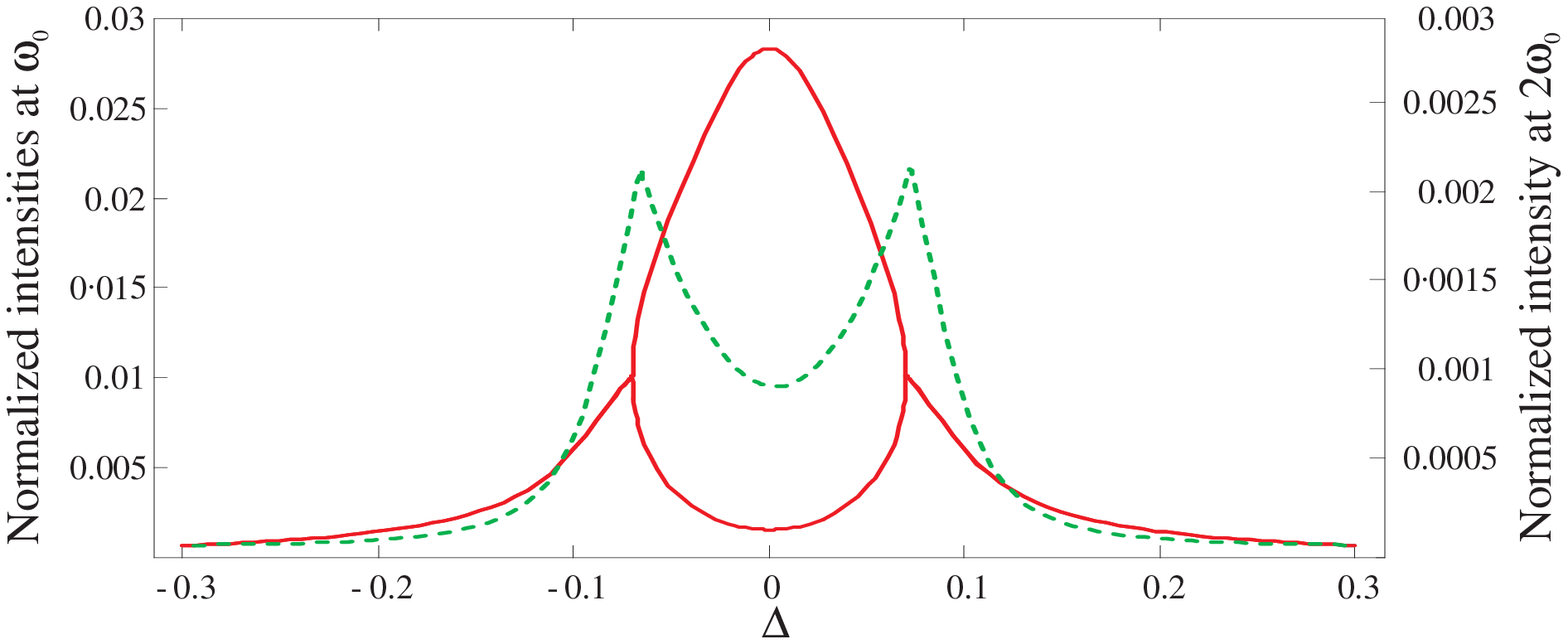}
\end{center}
\caption{\label{fig:theorie} Top : normalized intensities of the
fundamental frequency as a function of $I_{in}$ with $\Delta=0$.
Bottom : normalized intensities of the extraordinary and ordinary
fundamental frequency (continuous line) and harmonic frequency
fields (dashed line) as a function of $\Delta$ with
$I_{in}=0.001$. The other parameters are $\gamma_0=0.06$,
$\gamma^\prime=\gamma=0.11$, $\Delta_0=0$.}
\end{figure}

In the symmetric case, the intracavity intensity $I$ of both
fields at the fundamental frequency is a solution of a third
degree equation~:
\begin{equation}
\left(\gamma^{\prime2} +\Delta^2 + \frac{ I^2}{\gamma_0^2 +
\Delta_0^2}  + 2 \frac{\gamma \gamma_0 - \Delta
\Delta_0}{\gamma_0^2 + \Delta_0^2} I \right)I = \gamma |A_{in}|^2
\label{eq:solsym}
\end{equation}
In the dissymmetric case, $I_1$ and $I_2$ verify~:
\begin{eqnarray}
I_1 I_2 &=& (\gamma_0^2 + \Delta_0^2)(\gamma^{\prime2} +
\Delta^2) \label{eq:dissym}\\
 I_1 + I_2  &=& \frac{\gamma |A_{in}|^2}{\gamma^2 +
\Delta^2} - 2 (\gamma \gamma_0 - \Delta
\Delta_0)\label{eq:soldissym}
\end{eqnarray}
It can be shown by a linear stability analysis \cite{ou94} that
the symmetric solution is stable for~:
\begin{eqnarray}
I_{in} &<& I_{threshold}=
\frac{2\left(\gamma^{\prime2}+\Delta^2\right)}{\gamma} \times
\nonumber \\&& \left(\sqrt{\left(\gamma_0^2 +
\Delta_0^2\right)\left(\gamma^2 + \Delta^2\right)}+ \gamma
\gamma_0 -\Delta \Delta_0 \right) \label{eq:seuil}
\end{eqnarray}
while the dissymmetric solution is stable in the opposite case
($I>I_{threshold}$). The symmetry-breaking phenomenon can be
understood in the following way : above $I_{threshold}$, frequency
degenerate parametric oscillation takes place. The generated
subharmonic field can be shown \cite{ou94} to be polarized the
direction (DP) of figure (\ref{fig:ringcav}), \emph{i.e.}
orthogonally to the pump polarization (IP). The total field at
frequency $\omega_0$ is therefore no longer at 45$^\circ$ from the
crystal axes, leading to the symmetry breaking phenomenon.
Furthermore, because of the well-known $\pi$ phase indeterminacy
of the subharmonic field in the degenerate OPO, the vectorial sum
of the fields at frequency $\omega_0$ can take two different
values : as a result, the lower and upper values of $I_1$ and
$I_2$ can be taken by any of the two polarizations (1) or (2), and
the system may switch from one solution to the other. Equations
(\ref{eq:solsym}) and (\ref{eq:soldissym}) allow one to plot the
intracavity intensities as a function of the various parameters of
the system. The corresponding plots are shown in fig.
\ref{fig:theorie}.

The top graph shows a pitchfork bifurcation phenomenon as a
function of the input intensity : below a critical value, the
system is symmetric and above this value it is asymmetric. The
bottom graph gives the same phenomenon as a function of cavity
detunings. For large values of $|\Delta|$, the fundamental
frequency waves are far from resonance and $I_{threshold}$ is
large : the intracavity intensities are symmetric and follow the
usual cavity resonance shape. Under a certain value of $|\Delta|$,
the harmonic frequency intensity becomes larger than the OPO
oscillation threshold, parametric reconversion occurs and the
harmonic intensity decreases : this is the phenomenon of pump
depletion which occurs in standard OPOs. In our case, the
spontaneous symmetry breaking occurs~: the fundamental frequency
intensities are no longer equal as is the case in a standard
harmonic-pumped OPO or in a non-degenerate system.

\begin{figure}[h]
\centerline{\includegraphics[width=\columnwidth,clip=]{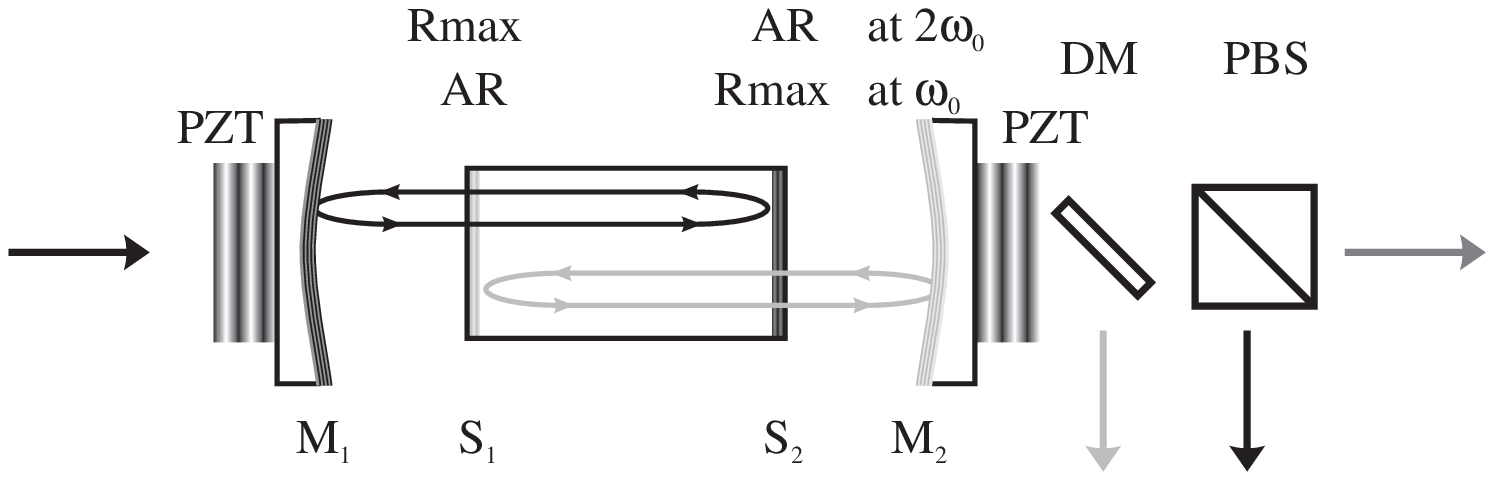}}
\caption{Principle of the dual-cavity. DM : dichroic mirror; PBS :
polarizing beamsplitter; PZT : piezo-electric ceramic.
\label{fig:dualcavity}}
\end{figure}

We will now describe the experimental set-up. As mentioned
previously, we need to be able to control independently $\Delta$
and $\Delta_0$. In order to achieve this, we use a dual-cavity
set-up (fig. \ref{fig:dualcavity}) : this also adds a degree of
freedom which allows the verification of relation
(\ref{eq:deltas}). The intensity reflection coefficient are shown
in table \ref{tab:characteristics}. The fundamental frequency beam
is produced by a Nd:YAG laser ($\lambda=1064~nm$, Lightwave
126-1024-700) which is mode-matched to the cavity and linearly
polarized at $45^\circ$ of a KTP crystal neutral axes (Cristal
Laser).

\begin{table}[h]
\caption{Intensity reflection coefficients.
\label{tab:characteristics} }
\begin{center}
\begin{tabular}{ccccc}
& $M_1$ & $S_1$ & $S_2$ & $M_2$\\ \hline \hline $R(1064~nm)$&
$95~\%$ &
$0.11~\%$ & $99.96~\%$ & $99.8~\%$ \\
$R(532~nm)$ & $>99.9~\%$ & $99.3~\%$ &  $5.25~\%$& $95~\%$ \\
\end{tabular}
\end{center}
\end{table}

The crystal is shared by the two cavities, one resonant for the
fundamental frequency (IR cavity), the other for the harmonic
frequency (green cavity). The green cavity length is kept
approximately fixed but not locked to a well defined value of the
detuning $\Delta_0$ while the IR cavity length is scanned via the
PZT ceramic. The intracavity intensities are monitored via the
small transmission of the green cavity and we are able to measure
independently the intensities along the crystal neutral axes.

\begin{figure}
  \centering
   \includegraphics[width=.75\columnwidth,clip=]{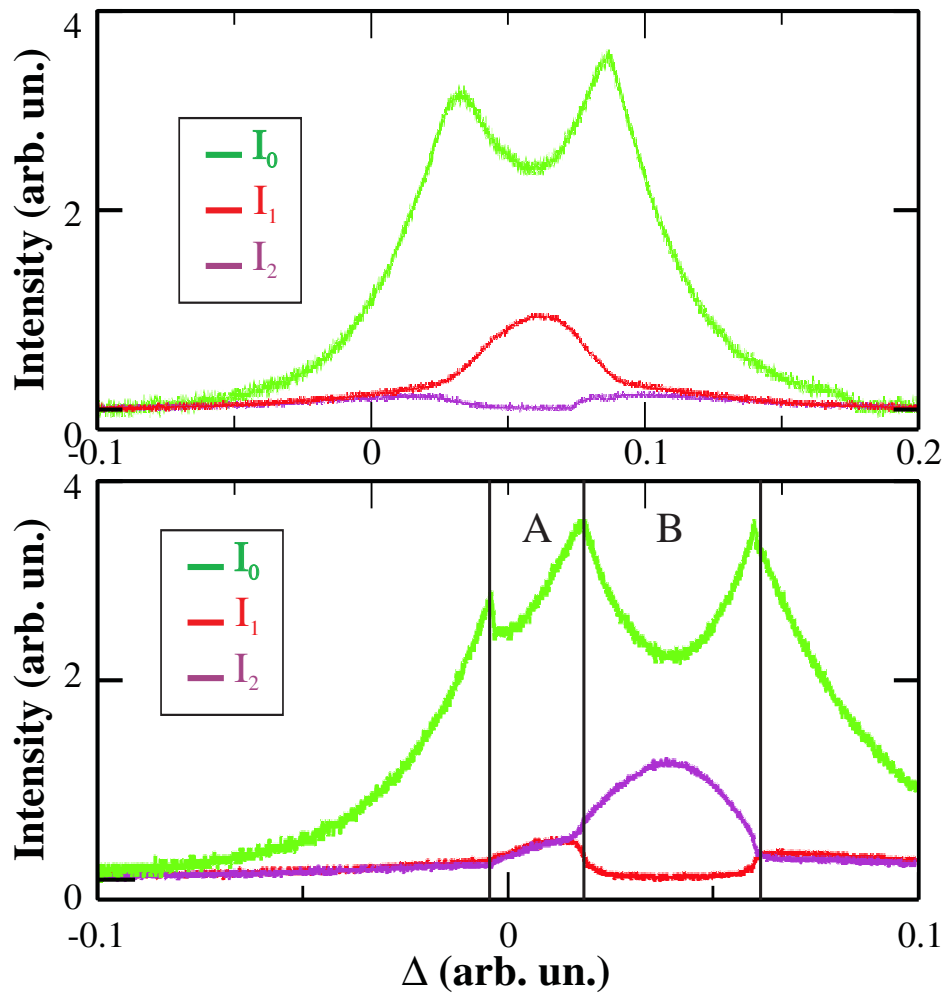}
  \caption{Experimental recordings of the intracavity intensities as a
  function of the cavity length scanned in time.}
  \label{fig:bifurcationsimple}
\end{figure}

A spontaneous symmetry breaking is observed for certain values of
the IR cavity length (fig. \ref{fig:bifurcationsimple}).
Experimental values of the various parameters governing the system
behavior vary from one recording to the other. In some recordings,
one observes that $I_2>I_1$ (fig. \ref{fig:bifurcationsimple},
top), in others that $I_2<I_1$ (B region of fig.
\ref{fig:bifurcationsimple}, bottom). In both cases, one observes
pump depletion as predicted on fig. \ref{fig:theorie}, bottom.
Pump depletion is also observed in the region A of fig.
\ref{fig:bifurcationsimple}, but without symmetry breaking between
$I_1$ and $I_2$ : this region of the parameter space corresponds
to the non frequency degenerate operation, which does not lead to
symmetry breaking.

 In conclusion, we have developed a dual-cavity set-up which
has allowed us to observe the spontaneous symmetry breaking
predicted in intracavity frequency generation with triple
resonance. To our knowledge, this is the first experimental
evidence for this behavior.

Laboratoire Kastler-Brossel, of the Ecole Normale Sup\'{e}rieure
and the Universit\'{e} Pierre et Marie Curie, is associated with
the Centre National de la Recherche Scientifique (UMR 8552).
Laboratoire Mat{\'e}riaux et Ph{\'e}nom{\`e}nes Quantiques is a F{\'e}d{\'e}ration de
Recherche, FR CNRS 2437. This work has been supported by
France-Telecom (project CTI n$^\circ$ 98-9.003), the European
Commission project QUICOV (IST-1999-13071) and ACI Photonique
(Minist\`ere de la Recherche).

\end{document}